\documentclass[aps,preprint,amssymb,eqsecnum,showpacs,tightenlines,nofootinbib]{revtex4}

\usepackage{amsmath}
\usepackage{amssymb}

\newcommand{\vect}[1]{\mathbf{#1}}

\newcommand{\Lag}{\mathcal{L}}

\newcommand{\eps}{\varepsilon}

\newcommand{\ket}[1]{\left|{#1}\right\rangle}
\newcommand{\braket}[2]{\left\langle{#1}|{#2}\right\rangle}
\newcommand{\A}{\mathcal{A}}
\newcommand{\comm}[2]{\left[{#1},{#2}\right]}

\newcommand{\tendsto}{\rightarrow}
\newcommand{\infinity}{\infty}

\newcommand{\comment}[1]{}
\newcommand{\beq}{\begin{equation}}
\newcommand{\eeq}{\end{equation}}
\newcommand{\beqa}{\begin{eqnarray}}
\newcommand{\eeqa}{\end{eqnarray}}

\begin{document}

\title{Quantum corrections to the Larmor radiation formula in scalar electrodynamics}

\author{A.~Higuchi}
\email{ah28@york.ac.uk}
\author{P.~J.~Walker}
\email{pjw120@york.ac.uk}
\affiliation{Department of Mathematics, University of York, Heslington, York YO10 5DD, United Kingdom}

\date{August 16, 2009}

\begin{abstract}
We use the semi-classical approximation in perturbative scalar quantum electrodynamics to calculate the quantum correction to  the Larmor radiation formula to first order in Planck's constant in the non-relativistic approximation, choosing the 
initial state of the charged particle to be a momentum eigenstate.  We calculate this
correction in two cases: in the first case the charged particle is accelerated by a time-dependent but space-independent
vector potential whereas in the second case it is accelerated by a time-independent vector 
potential which is a function of one spatial coordinate.  We find that the corrections in these two cases are different even for a charged particle with the same classical motion.  The correction in each case turns out to be non-local in time in contrast to the classical approximation. 
\end{abstract}


\maketitle

\section{Introduction}

A well-known result in classical electrodynamics, discovered during the burst of activity in the late nineteenth century, is that an accelerated charge emits radiation.  In particular, the formula which gives the amount of energy radiated by 
the charge was found by Larmor in this period.  The relativistic generalization of this formula is
\beq
E_{\rm em}^{(0)} = -\frac{e^2}{6\pi c^3} \int dt \frac{d^2x^\mu}{d\tau^2} \frac{d^2x_\mu}{d\tau^2},  \label{Larmor}
\eeq
where $e$ is the charge of the particle and $c$ is the speed of light. Here and below the metric signature is $+---$ and
$\tau$ is the proper time along the world line of the particle, $x^\mu(\tau)$, with $x^0=ct$.  
(See Ref.~\cite{Jackson}, Sec.~14.2, for a derivation of this result.)

Since classical electrodynamics is an approximation to quantum electrodynamics (QED), one 
expects that the Larmor formula should be reproduced in the latter theory in the limit $\hbar \to 0$ (at
order $e^2$).  Indeed it has been shown that this formula is recovered 
in QED for a scalar charged particle moving on a straight line
in the limit $\hbar\to 0$~\cite{HiguchiMartin05}. Furthermore
it has been shown~\cite{HiguchiMartin06a,HiguchiMartin06b} that the
Lorentz-Dirac radiation-reaction force~\cite{Abraham,Lorentz,Dirac} is obtained in the limit $\hbar\to 0$ in QED for a scalar charged particle in
three-dimensional motion under the influence of a vector potential depending only on one spacetime 
coordinate. (For other approaches for studying the Lorentz-Dirac force in the context of QED, see Refs.~\cite{MS,Tsyt,FordOConnell,OConnell,beilok}.)
This work indirectly shows that the Larmor formula is reproduced in the limit $\hbar\to 0$ 
for a charged scalar particle in three-dimensional motion under the conditions specified because the Lorentz-Dirac force
and energy-momentum conservation imply the Larmor formula.

Although the Larmor formula correctly gives the amount of energy emitted as radiation in the limit $\hbar\to 0$, it is clearly not exact. For example, in Ref.~\cite{NoSaYa} a model with a scalar charged particle which is soluble to
order $e^2$ in QED was studied and the exact result for the energy emitted was shown to differ from the Larmor formula.
It will be interesting, therefore, to estimate the correction of order $\hbar$ to the Larmor formula for general motion of the charged particle.  The purpose of this paper is to carry out this task in the simple setting used in Refs.~\cite{HiguchiMartin05,HiguchiMartin06a,HiguchiMartin06b} where the scalar particle is accelerated by a vector potential that depend only on one spacetime coordinate under the additional condition that the initial state of the charged particle 
is a momentum eigenstate.

One might hope that there would be
a universal expression for this correction which depended only on the motion of the corresponding classical particle, but we find that the correction 
depends on how the particle is accelerated.
For this reason we calculate the quantum correction to the Larmor formula at order $\hbar$ in two cases: in the first case the charged particle is accelerated by a time-dependent but space-independent vector potential whereas in the second case it is accelerated by a time-independent vector potential which is a function of one spatial coordinate.  We also use the non-relativistic approximation because a fully relativistic calculation would be too complicated for the purpose of
this paper, which is to show how the quantum correction to the Larmor formula can be found in simple examples.

The rest of the paper is organized as follows.  In Sec.~\ref{sec:Larmor}, we show directly that the Larmor formula is reproduced in scalar QED in the limit $\hbar \tendsto 0$.  We then proceed in Secs.~\ref{sec:Corrections-time} and
\ref{sec:Corrections-space} to calculate the correction to this formula at order $\hbar$ in the two cases mentioned above.
Finally, in Sec.~\ref{sec:Conclusions} we provide a summary and concluding remarks.  
Throughout this paper we retain $\hbar$ explicitly but let $c=1$ except where it is convenient not to do so.

\section{The Larmor formula in QED}\label{sec:Larmor}

In this section we derive the Larmor formula from QED for a charged scalar particle accelerated by 
a vector potential $V^\mu$ which depends
only on $t$.  We follow Refs.~\cite{HiguchiMartin05,HiguchiMartin06a} closely.  (The derivation for the case with a potential
which depends on one space coordinate will not be presented, but it is very similar to the case treated here.) 
We assume that the variation in $V^\mu(t)$ occurs only over a bounded interval $[-T,T]$, $T>0$. We let $V^\mu(t)=0$ for
$t < -T$ without loss of generality and $V^\mu(t)$ for $t>T$ be a constant which is not necessarily zero.\footnote{In 
Refs.~\cite{HiguchiMartin05,HiguchiMartin06a}
the convention was slightly different in that
$V^\mu$ was chosen to satisfy $V^\mu(t)=0$ for positive $t$.} 
We also use a gauge transformation to impose the condition 
$V_0(t) = 0$ for all $t$.

The Lagrangian density of our model is
\beq
\Lag = \left[\left(D_\mu + ieA_\mu\right)\phi\right]^\dagger \left[\left(D^\mu +ieA^\mu\right)\phi\right] - \frac{m^2}{\hbar^2} \phi^\dagger \phi - \frac{1}{4} F_{\mu\nu} F^{\mu\nu} - \frac{1}{2}(\partial_\mu A^\mu)^2,
\eeq
where $D_\mu \equiv \partial_\mu + \frac{i}{\hbar} V_\mu$.  We have adopted the Feynman gauge, in which the non-interacting field equation, i.e. the field equation with $e=0$, 
for $A_\mu$ is $\partial_\nu \partial^\nu A^\mu = 0$. We can therefore expand it 
in terms of momentum modes,
\beq
A_\mu(x) = \int \frac{d^3 \vect{k}}{2k(2\pi)^3} \left[a_\mu(\vect{k}) e^{-ik \cdot x} + {a_\mu}^\dagger(\vect{k}) e^{ik \cdot x}\right],
\eeq
where $k = \left|\vect{k}\right|$.  The operators $a_\mu(\vect{k})$ and ${a_\mu}^\dagger(\vect{k})$ obey the usual commutation relations,
\beq
\comm{a_\mu(\vect{k})}{{a_\nu}^\dagger(\vect{k}')} = - 2\hbar k (2\pi)^3 g_{\mu\nu} \delta^3(\vect{k} - \vect{k}').
\label{photoncomm}
\eeq
We can use the Fourier expansion for the scalar field as well.  
Thus we write
\beq
\phi(x) = \hbar \int \frac{d^3 \vect{p}}{2p_0(2\pi\hbar)^3} \left[A(\vect{p}) \Phi_{\vect{p}}(x) + B^\dagger(\vect{p}) \overline{\Phi}_{\vect{p}}^*(x) \right].
\eeq
The mode functions $\Phi_{\vect{p}}(x)$ 
are different from the standard `free' mode functions, 
$e^{-ip\cdot x/\hbar}$. (We do not need to consider the anti-particle modes $\overline{\Phi}_{\vect{p}}(x)$  though their relation to $\Phi_{\vect{p}}(x)$ is very simple.)
This is because the equation of motion for the scalar field with $e=0$ is not the free field equation, but rather,
\beq
\left(\hbar^2 D_\mu D^\mu + m^2\right) \Phi_{\vect{p}}(x) = 0.
\eeq
Since the potential $V_\mu(t)$ depends only on $t$, these mode functions can be written in the following form:
\begin{equation}\label{eq:wkb}
\Phi_{\vect{p}}(x) = \sqrt{p_0} \phi_{\vect{p}}(t) \exp\left(\frac{i}{\hbar}\vect{p}\cdot\vect{x}\right),
\end{equation}
where $p_0 = \sqrt{|\vect{p}|^2 + m^2}$.  Since we are interested in the limit $\hbar\to 0$,  
we use the WKB approximation, which gives 
\beq
\phi_{\vect{p}}(t) = \frac{1}{\sqrt{\sigma_{\vect{p}}(t)}} \exp{\left[-\frac{i}{\hbar}\int_0^t \sigma_{\vect{p}}(\zeta)d\zeta\right]}\psi_\mathbf{p}(t), \label{WKB}
\eeq
where 
\beq
\sigma_{\vect{p}}(t) \equiv \sqrt{|\vect{p}-\vect{V}(t)|^2+m^2} \label{kinetic}
\eeq
is the kinetic energy of a scalar particle with momentum $\vect{p}$.  The function $\psi_\vect{p}(t)$ contains
the corrections of higher order in $\hbar$, i.e.
\beq
\psi_\mathbf{p}(t) = 1 + i\hbar g_\vect{p}(t) +O(\hbar^2). 
\eeq
It can readily be shown that $g_\vect{p}(t)$ is real.
The non-trivial commutation relations among annihilation and creation operators are
\beq
\left[ A(\vect{p}),A^\dagger(\vect{p}')\right]
= \left[B(\vect{p}),B^\dagger(\vect{p}')\right] = 2p_0(2\pi\hbar)^3\delta^3(\vect{p}-\vect{p}').
\label{particlecomm}
\eeq
The operators $A^\dagger(\vect{p})$ and $B^\dagger(\vect{p})$ create a particle and an anti-particle, respectively.

The initial state with one charged scalar particle and no photon can be given in general as
\beq
\ket{i} = \int \frac{d^3 \vect{p}}{\sqrt{2p_0}(2\pi\hbar)^3} f(\vect{p}) A^\dagger(\vect{p}) \ket{0}.
\eeq
This state is normalized so that $\braket{i}{i} = 1$. This condition implies
\beq
\int \frac{d^3 \vect{p}}{(2\pi\hbar)^3} \left|f(\vect{p})\right|^2 = 1.
\eeq
It is sufficient to assume that the function $f(\vect{p})$ is peaked about a given momentum with width of order $\hbar$
to derive the Larmor formula.
However, this assumption will not be sufficient when we come to consider its quantum correction.  
For this reason we assume
that $f(\vect{p})$ is sharply peaked with an arbitrary accuracy and take the limit such that $|f(\vect{p})|^2$ is
proportional to a delta-function at an appropriate stage.  This procedure amounts to the condition that the initial state
is a momentum eigenstate.

An initial state with one charged particle evolves in general to order $e^2$ as
\beq
A^\dagger (\vect{p}) \ket{0} \mapsto [1+i\hbar^{-1}\mathcal{F}(\vect{p})]A^\dagger(\vect{p})\ket{0} + \frac{i}{\hbar}\int \frac{d^3 \vect{k}}{2k(2\pi)^3} \A^\mu(\vect{p},\vect{k}) {a_\mu}^\dagger(\vect{k}) A^\dagger(\vect{P}) \ket{0},
\eeq
where $\vect{P} = \vect{p} - \hbar \vect{k}$ is the out-going momentum of the scalar particle when a photon is emitted,
$\A^\mu(\vect{p},\vect{k})$ is the amplitude for the emission of one photon, and
$\mathcal{F}(\vect{p})$ is the forward-scattering amplitude, which plays no role in this paper.
Thus, the initial state $\ket{i}$ evolves to
\beq
\ket{f} = \ket{f_0} + \ket{f_1},
\eeq
where
\begin{align}
\ket{f_0} &= \int \frac{d^3 \vect{p}}{\sqrt{2p_0}(2\pi\hbar)^3} \left[1+i\hbar^{-1}\mathcal{F}(\vect{p})\right]f(\vect{p}) A^\dagger(\vect{p})\ket{0},\\
\ket{f_1} & = \frac{i}{\hbar} \int \frac{d^3 \vect{p}}{\sqrt{2p_0}(2\pi\hbar)^3} \int \frac{d^3 \vect{k}}{2k(2\pi)^3} f(\vect{p}) \A^\mu(\vect{p},\vect{k}) {a_\mu}^\dagger(\vect{k}) A^\dagger(\vect{P}) \ket{0}. \label{eq:final-state}
\end{align}
The emission probability in the limit where $f(\mathbf{p})$ is arbitrarily 
sharply peaked can be found using the commutation relations
(\ref{photoncomm}) and (\ref{particlecomm}) as
\beqa
\Gamma & = & \langle f_1\,|\,f_1\rangle \nonumber \\
& = & \frac{1}{\hbar}\int \frac{d^3\vect{k}}{2k(2\pi)^3}\frac{P_0}{p_0}\left|\frac{\partial\vect{P}}{\partial\vect{p}}\right|^{-1}
|\mathcal{A}(\vect{p},\vect{k})|^2, \label{Gamma}
\eeqa
where $|\mathcal{A}(\vect{p},\vect{k})|^2 \equiv - \mathcal{A}_\mu^*(\vect{p},\vect{k})\mathcal{A}^\mu(\vect{p},\vect{k})$ and
where
\beq
\frac{\partial\vect{P}}{\partial\vect{p}} \equiv {\rm det}\left(\frac{\partial P^i}{\partial p^j}\right)
\label{Jacobian}
\eeq
is the Jacobian determinant.  The momentum $\mathbf{p}$ 
is now the peak value of the momentum distribution of the initial state.
The energy emitted is obtained by multiplying the integrand in Eq.~(\ref{Gamma}) by the photon energy, $\hbar k$. 
We have $\partial\vect{P}/\partial\vect{p}=1$ because $\vect{P}=\vect{p}-\hbar\vect{k}$.
Hence, the $4$-momentum of the radiation emitted is
\beq
\mathcal{P}^\mu = \int \frac{d^3\vect{k}}{16\pi^3}\frac{P_0}{p_0}\,n^\mu|\mathcal{A}(\vect{p},\vect{k})|^2,  \label{Pmu}
\eeq
where $n^\mu \equiv k^\mu/k$.
It can be shown~\cite{HiguchiMartin06a,HiguchiMartin06b} that
\begin{equation}\label{eq:em-amp}
\A_\mu(\vect{p},\vect{k}) = -ie\hbar \int \frac{d^3 \vect{p}'}{2p_0'(2\pi\hbar)^3} \int d^4 x e^{ik\cdot x} \left[\Phi^*_{\vect{p}'}(x) D_\mu \Phi_{\vect{p}}(x) - (D_\mu \Phi_{\vect{p}'}(x))^* \Phi_{\vect{p}}(x)\right].
\end{equation}
Since $\Phi_{\vect{p}}(x) = \sqrt{p_0} \phi_{\vect{p}}(t) e^{i\vect{p}\cdot\vect{x}/\hbar}$, the
exponential factors in the integrand of Eq.~(\ref{eq:em-amp}) result in 
$(2\pi\hbar)^3\delta^3(\vect{p} - \hbar\vect{k} - \vect{p}')$ upon integration over $\vect{x}$.  Thus, we find
\begin{align}
\A_i(\vect{p},\vect{k}) &= -\frac{e}{2} \sqrt{\frac{p_0}{P_0}}
\int dt \, e^{ikt} \phi_{\vect{P}}^*(t) \phi_{\vect{p}}(t) \left[p_i + P_i - 2 V_i(t)\right], \label{sp}\\
\A_0(\vect{p},\vect{k}) &= -\frac{ie\hbar}{2}\sqrt{\frac{p_0}{P_0}} \int dt \, e^{ikt} \left[\phi_{\vect{P}}^*(t) \frac{d\phi_{\vect{p}}(t)}{dt} - \frac{d\phi_{\vect{P}}^*(t)}{dt} \phi_{\vect{p}}(t)\right]. \label{ti}
\end{align}

Now we use the WKB approximation (\ref{WKB}) and find
\beq
\phi_{\vect{P}}^*(t)\phi_{\vect{p}}(t) = \frac{1}{\sqrt{\sigma_{\vect{P}}(t)\sigma_{\vect{p}}(t)}}
\exp\left\{-\frac{i}{\hbar} \int_0^t \left[\sigma_{\vect{p}}(\zeta) - \sigma_{\vect{P}}(\zeta)\right]d\zeta\right\}
\psi_\vect{P}^*(t)\psi_\vect{p}(t). \label{product}
\eeq
To lowest order in $\hbar$ we have
\begin{equation}
-\frac{i}{\hbar} \int_0^t [\sigma_{\vect{p}}(\zeta) - \sigma_{\vect{P}}(\zeta)]d\zeta \approx -i\vect{k}\cdot \int_0^t \frac{\vect{p}-\vect{V}(\zeta)}{\sigma_{\vect{p}}(\zeta)}d\zeta,  \label{approx}
\end{equation}
where the relation $\vect{P}=\vect{p}-\hbar \vect{k}$ has been used.  If $\vect{x}(t)$ is the position of a classical particle
corresponding to the state $A^\dagger(\vect{p})|0\rangle$, i.e.\ 
with momentum $\vect{p}$ under the influence of the vector potential $\vect{V}(t)$, then
\begin{align}
m\frac{d\vect{x}}{d\tau} &= \vect{p} - \vect{V}(t), \label{space}\\
m\frac{dt}{d\tau} &= \sigma_{\vect{p}}(t). \label{time}
\end{align}
These relations imply $[\vect{p}-\vect{V}(t)]/\sigma_\vect{p}(t) \approx d\vect{x}/dt$ to lowest order in $\hbar$.  
Using this approximation in Eq.~(\ref{approx}) and substituting the result into Eq.~(\ref{product})
and requiring $\vect{x}(0)=0$, we find to lowest order in $\hbar$ that
\beq
\phi_{\vect{P}}^*(t)\phi_{\vect{p}}(t) \approx \frac{1}{\sigma_\vect{p}(t)}e^{-i\vect{k}\cdot\vect{x}}. \label{pro}
\eeq
Also it can readily be shown that
\beq
i\hbar \frac{d\phi_{\vect{p}}(t)}{dt} \approx \sigma_{\vect{p}}(t) \phi_{\vect{p}}(t).  \label{temporal}
\eeq
Substituting Eqs.~(\ref{pro}) and (\ref{temporal}) into Eqs.~(\ref{sp}) and (\ref{ti}), and using 
Eqs.~(\ref{space}) and (\ref{time}), we find
\begin{align}
\A^0(\vect{p},\vect{k}) &= -e \int dt e^{ik\cdot x}\\
\A^i(\vect{p},\vect{k}) &= -e \int dt \frac{dx^i}{dt} e^{ik\cdot x},
\end{align}
which can be combined as
\beq
\A^\mu(\vect{p},\vect{k}) = -e \int d\xi \frac{dx^\mu}{d\xi} e^{ik\xi}, \label{ill}
\eeq
where $\xi \equiv n\cdot x$.  

Eq.~(\ref{ill}) is ill-defined because $dx^\mu/d\xi$ is finite for arbitrarily large values of $|\xi|$.
We therefore introduce a compactly supported cut-off factor, $\chi(a\xi)$, $0<a \leq 1$, 
which is 1 on a compact interval including the region where the acceleration takes place, and smoothly varies between $0$ and $1$. Then the emission amplitude becomes
\beqa
\A^\mu(\vect{p},\vect{k}) & = & - e\int d\xi \frac{dx^\mu}{d\xi}\chi(a\xi)e^{ik\xi} \label{em-amp} \\
& = & -\frac{ie}{k} \int d\xi \left[\frac{d^2x^\mu}{d\xi^2} \chi(a\xi) 
+ a\frac{dx^\mu}{d\xi}\chi'(a\xi)\right] e^{ik\xi}.  \label{eq:em-amp-final}
\eeqa
By substituting this equation into Eq.~(\ref{Pmu}) and taking the limit $a\to 0$, we find the $4$-momentum of the radiation emitted to lowest order in $\hbar$ and $e$ as
\beq
\mathcal{P}^\mu = -\frac{e^2}{16\pi^2} \int d\Omega\, \int d\xi n^\mu \frac{d^2 x^\nu}{d\xi^2} \frac{d^2 x_\nu}{d\xi^2},
\eeq
where $d\Omega$ is the solid-angle for the unit vector $\vect{n} = \vect{k}/k$.
We convert the $\xi$-derivative to the $t$-derivative by using the formula
$d\xi/dt = n_\mu dx^\mu/dt$ as
\beq
\frac{d^2x^\mu}{d\xi^2} = \left(\frac{dt}{d\xi}\right)^3\left(\frac{d\xi}{dt}\frac{d^2x^\mu}{dt^2} - \frac{d^2\xi}{dt^2}\frac{dx^\mu}{dt}\right).
\eeq
The result is
\beq
\mathcal{P}^\mu = -\frac{e^2}{16\pi^2} \int dt \int d\Omega \dot{\xi}^{-5} n^\mu n_\sigma n_\rho \left[\dot{x}^\sigma \dot{x}^\rho \ddot{x}^\nu \ddot{x}_\nu - 2 \dot{x}^\sigma \ddot{x}^\rho \ddot{x}^\nu \dot{x}_\nu + \ddot{x}^\sigma \ddot{x}^\rho \dot{x}^\nu \dot{x}_\nu\right],
\eeq
where the dot indicates the $t$-derivative.  The integration over the solid angle can be carried out by 
using (see Ref.~\cite{HiguchiMartin06a})
\beq
\int d\Omega\, \dot{\xi}^{-5} n^\mu n_\sigma n_\rho = \frac{4}{3}\pi \left[6\gamma^8 \dot{x}^\mu \dot{x}_\sigma \dot{x}_\rho - \gamma^6 \left(\delta^\mu_\sigma \dot{x}_\rho + \dot{x}^\mu g_{\sigma\rho} + \delta^\mu_\rho \dot{x}_\sigma\right) \right],
\eeq
where $\gamma \equiv dt/d\tau = (\dot{x}^\mu\dot{x}_\mu)^{-1/2}$.
Thus we obtain
\beq
\mathcal{P}^\mu= -\frac{e^2}{6\pi} \int dt \, \dot{x}^\mu \left[\gamma^4\ddot{x}\cdot\ddot{x} - 
\gamma^6(\dot{x} \cdot \ddot{x})^2 \right].
\eeq
By converting the $t$-derivative to the $\tau$-derivative, we
find
\beq
\mathcal{P}^\mu = -\frac{e^2}{6\pi c^4} \int d\tau \frac{dx^\mu}{d\tau} \frac{d^2x^\nu}{d\tau^2} \frac{d^2x_\nu}{d\tau^2},
\label{Larmor4}
\eeq
which is a well-known result in classical electrodynamics, and
the component $\mathcal{P}^0c$ gives the Larmor formula (\ref{Larmor}).

\section{Quantum correction with time-dependent vector potential}\label{sec:Corrections-time}

In this section we calculate the correction to the Larmor formula at order $\hbar$ to lowest order in the non-relativistic
approximation in the case where the charged scalar particle is accelerated by a time-dependent but space-independent 
vector potential.

{}From Eq.~(\ref{Pmu}) we find the energy emitted as
\beq
E_{\rm em} = 
-\int \frac{d^3\vect{k}}{16\pi^3}\frac{P_0}{p_0} \mathcal{A}_\mu^*(\vect{p},\vect{k}) 
\mathcal{A}^\mu(\vect{p},\vect{k}), \label{emitted}
\eeq
where $\mathcal{A}_i(\vect{p},\vect{k})$ and $\mathcal{A}_0(\vect{p},\vect{k})$ are given by
Eqs.~(\ref{sp}) and (\ref{ti}), respectively.
By substituting the WKB expression for $\phi_\vect{p}(t)$ given by Eq.~(\ref{WKB}) into these equations we find
\begin{align}
\A^0(\vect{p},\vect{k}) = &-\frac{ec}{2} \sqrt{\frac{p_0}{P_0}} \int dt e^{i\omega t} \frac{1}{\sqrt{\sigma_{\vect{p}}(t) \sigma_{\vect{P}}(t)}}\nonumber\\
	&\times\left\{\sigma_{\vect{p}}(t) + \sigma_{\vect{P}}(t) + i\hbar c\left[\frac{\sigma_{\vect{P}}'(t)}{2\sigma_{\vect{P}}(t)} -\frac{\sigma_{\vect{p}}'(t)}{2\sigma_{\vect{p}}(t)} + \frac{\psi_{\vect{p}}'(t)}{\psi_{\vect{p}}(t)}
	-  \frac{\psi_{\vect{P}}^{*\prime}(t)}{\psi_{\vect{P}}^*(t)}\right]\right\}\nonumber\\
	&\times\exp{\left[-\frac{ic}{\hbar} \int_0^t\left(\sigma_{\vect{p}}(\zeta) - \sigma_{\vect{P}}(\zeta)\right)d\zeta\right]}\psi^*_{\vect{P}}(t)\psi_{\vect{p}}(t), \label{eq:A_0(t)}\\
\A^i(\vect{p},\vect{k}) = & - \frac{ec}{2} \sqrt{\frac{p_0}{P_0}} \int dt e^{i\omega t} \frac{1}{\sqrt{\sigma_{\vect{p}}(t) \sigma_{\vect{P}}(t)}}\left[p^i - V^i(t) + P^i - V^i(t)\right]\nonumber\\
	&\times\exp{\left\{-\frac{ic}{\hbar} \int_0^t\left[\sigma_{\vect{p}}(\zeta) - \sigma_{\vect{P}}(\zeta)\right]d\zeta\right\}}
	\psi^*_{\vect{P}}(t)\psi_{\vect{p}}(t), \label{eq:A_i(t)}
\end{align}
where $\sigma_\vect{p}=\sqrt{|\vect{p}-\vect{V}|^2+m^2c^2}$ and $\omega = kc$.  
We have restored factors of $c$ by dimensional analysis,
anticipating the use of the non-relativistic approximation.
It is straightforward to calculate the amplitude to order $\hbar$ using $\vect{P}=\vect{p}-\hbar\vect{k}$.  The result is
\begin{align}
\A^0(\vect{p},\vect{k}) = &-ec \sqrt{\frac{p_0}{P_0}} \int dt e^{i\omega t-i\vect{k}\cdot \vect{x}} 
\exp\left\{ i\frac{\hbar c}{2}\int_0^t \left[\frac{k^2}{\sigma_\vect{p}(t)} - \frac{(\vect{k}\cdot\dot{\vect{x}})^2}{\sigma_\vect{p}(t)c^2}\right]d\zeta\right\}, \label{A0}\\
\A^i(\vect{p},\vect{k}) = & - ec \sqrt{\frac{p_0}{P_0}} \int dt e^{i\omega t-i\vect{k}\cdot\vect{x}} \
\left\{ \frac{\dot{x}^i}{c} - \frac{\hbar}{2\sigma_\vect{p}(t)}\left[ k^i - \frac{\dot{x}^i(\vect{k}\cdot\dot{\vect{x}})}{c^2}\right]\right\}\nonumber \\
	&\times\exp{\left\{i\frac{\hbar c}{2}\int_0^t \left[\frac{k^2}{\sigma_\vect{p}(\zeta)} - \frac{(\vect{k}\cdot\dot{\vect{x}}(\zeta))^2}{\sigma_\vect{p}(\zeta)c^2}\right]d\zeta\right\}}. \label{Ai}
\end{align}
Note in particular that there is no contribution from the factor 
$\psi^*_\vect{P}(t)\psi_\vect{p}(t)\approx 1 + i\hbar(g_\vect{p}(t)-g_\vect{P}(t))$ at order $\hbar$.

One could write down a formal expression for the expected amount of 
energy emitted to order $\hbar$ by substituting these formulas into
Eq.~(\ref{emitted}).  Instead of doing so, we use the non-relativistic approximation 
in order to find an expression in terms of the classical trajectory of the particle in closed form.
We calculate the correction from the exponential factor common to both $\A^i$ and $A^0$ and that
from the additional term in $\A^i$ separately and add them up.

Denoting the correction due to the exponential factor by $\Delta E_1$, we have
\beqa
\Delta E_1
& = & \frac{ie^2\hbar}{32\pi^3 c^3}
\int d\Omega \int_0^\infty d\omega \omega^4\int dt dt'\,e^{i\omega(t-t')-i\omega\vect{n}\cdot
\left[\vect{x}(t)-\vect{x}(t')\right]/c}\nonumber \\
&& \times  \left[\dot{\vect{x}}(t)\cdot \dot{\vect{x}}(t')-c^2\right]
\int_{t'}^{t} \left[\frac{1}{\sigma_\vect{p}(\zeta)c} - \frac{(\vect{n}\cdot\dot{\vect{x}}(\zeta))^2}{\sigma_\vect{p}(\zeta)c^3}\right]d\zeta.
\eeqa
We use the non-relativistic approximation to order $c^{-5}$.  Thus, we expand the factor $e^{-i\omega\vect{n}\cdot \left[\vect{x}(t')-\vect{x}(t)\right]/c}$ with respect to $\omega/c$ to
order $c^{-2}$. (Notice that $\sigma_\vect{p}(t) \approx mc$ to lowest order in $c^{-1}$.)
Then we integrate over $\omega$, regularizing the integral by changing $e^{i\omega(t-t')}$ to 
$e^{i\omega(t-t'+i\eps)}$ and using the formula
\beq
\int_0^\infinity \omega^n e^{i\omega(t-t'+i\eps)} d\omega  =  i^{n+1} 
\frac{\partial^n\ }{\partial t^{\prime n}} \frac{1}{t-t' +i\eps}.
\eeq
Thus, we obtain
\beqa
\Delta E_1
& = & -\frac{e^2\hbar}{32\pi^3 c^3}
\int d\Omega\int dt dt'\left[ \frac{4!}{(t-t'+i\eps)^5} + \frac{6!\left\{\vect{n}\cdot\left[\vect{x}(t')-\vect{x}(t)\right]\right\}^2}{2(t-t'+i\eps)^7c^2}\right] \nonumber \\
&& \times  \left[\dot{\vect{x}}(t)\cdot \dot{\vect{x}}(t')-c^2\right]
\int_{t'}^{t} \left[\frac{1}{\sigma_\vect{p}(\zeta)c} - \frac{(\vect{n}\cdot\dot{\vect{x}}(\zeta))^2}{mc^4}\right]d\zeta.
\eeqa
This integral is ill-defined since the integrand remains finite if we let $|t+t'|$ be arbitrarily large 
while keeping $t-t'$ finite.  For this
reason we insert a cut-off factor $\chi(at)\chi(at')$, $0< a\leq 1$, such that $\chi(at)$ is smooth and compactly supported, 
and that $\chi(at)=1$ for $t \in [-T,T]$, i.e. while $V^\mu(t)$ is not constant. Then, we find that this integral is the sum of terms of the form $A_1^{(1)}$ and $A_1^{(3)}$ as defined in Eq.~(\ref{A1}).  Therefore, as is shown in Appendix~\ref{app:cut-off}, we can formally integrate by parts with respect
to $t$ and $t'$ to reduce the power of $t'-t+i\eps$ in the denominator.  Then we find
\beq
-c^2\int dt dt' \frac{4!}{(t-t'+i\eps)^5}\int_{t'}^{t} \left[\frac{1}{\sigma_\vect{p}(\zeta)c} - \frac{(\vect{n}\cdot\dot{\vect{x}}(\zeta))^2}{mc^4}\right]d\zeta = 0  \label{1}
\eeq
by integrating by parts with respect to $t$ and $t'$.  This means that, to find $\Delta E_1$ to order $c^{-5}$, we 
can let
\beq
\int_{t'}^{t} \left[\frac{1}{\sigma_\vect{p}(\zeta)c} - \frac{(\vect{n}\cdot\dot{\vect{x}}(\zeta))^2}{mc^4}\right]d\zeta
\approx \frac{1}{mc^2}(t-t'). \label{2}
\eeq
Hence we have
\beq
\Delta E_1
=  - \frac{e^2\hbar}{8\pi^3 mc^5}
\int d\Omega\int dt dt'\left[ \frac{3!\dot{\vect{x}}(t)\cdot \dot{\vect{x}}(t')}{(t-t'+i\eps)^4} - \frac{3\cdot 5!\left\{\vect{n}\cdot\left[\vect{x}(t')-\vect{x}(t)\right]\right\}^2}{4(t-t'+i\eps)^6}\right].
\eeq
Integrating the second term by parts with respects to $t$ and $t'$ and carrying out the $\vect{n}$-integration, we find
\beqa
\Delta E_1 & = & -\frac{e^2\hbar}{4\pi^2 mc^5}\int dt dt' \frac{3!\dot{\vect{x}}(t)\cdot \dot{\vect{x}}(t')}{(t-t'+i\eps)^4}
\nonumber \\
& = & -\frac{e^2\hbar}{8\pi^2 mc^5}\int dtdt'
\dot{\vect{x}}(t)\cdot\dot{\vect{x}}(t')
\left(\frac{\partial^3\ }{\partial t^2\partial t'} - \frac{\partial^3\ }{\partial t^{\prime 2}\partial t}\right)
\frac{1}{t-t'+i\eps}.
\eeqa
By integrating by parts, we find
\beq
\Delta E_1 = \frac{e^2 \hbar}{8\pi^2mc^5} \int dt \, dt' \left(\frac{d^3\vect{x}}{dt^3} \cdot \frac{d^2\vect{x}'}{dt'^2} - \frac{d^2\vect{x}}{dt^2} \cdot \frac{d^3\vect{x}'}{dt'^3}\right) \frac{1}{t-t'}.  \label{deltaE1}
\eeq

We move now to the correction which comes from the multiplicative factor in $\A^i(\vect{p},\vect{k})$.   
Since we only need this quantity to order $c^{-2}$, we find from Eq.~(\ref{Ai})
\beq
\A^i(\vect{p},\vect{k})|_{\rm non-ex}
\approx - e \sqrt{\frac{p_0}{P_0}} \int dt e^{i\omega t-i\omega \vect{n}\cdot\vect{x}/c}
\left[ \dot{x}^i(t) - \frac{\hbar \omega n^i}{2mc}\right],
\eeq
where we have dropped the correction to the exponential factor.  We find the corresponding correction in the Larmor formula
by substituting this formula in Eq.~(\ref{emitted}) as
\beq
\Delta E_2 = -\frac{e^2 \hbar}{32\pi^3 mc^4} \int d\Omega \int_0^\infty d\omega\omega^3 \int \, dt \, dt'
e^{i\omega(t-t')- i\omega \vect{n}\cdot (\vect{x}-\vect{x}')/c} \vect{n}\cdot \left(\dot{\vect{x}}' + \dot{\vect{x}}\right),
\eeq
where we have defined $x^i \equiv x^i(t)$ and $x^{\prime i} \equiv x^i(t')$. By expanding the factor
$e^{-i\omega \vect{n}\cdot(\vect{x}-\vect{x}')/c}$ to first order in $\omega/c$ and integrating over $\vect{n}$ and $\omega$
we find
\begin{equation}
\Delta E_2 = \frac{e^2 \hbar}{24\pi^2mc^5} \int dt \, dt'\frac{4!(\vect{x}'-\vect{x})\cdot(\dot{\vect{x}}+\dot{\vect{x}}')}
{(t-t'+i\eps)^5}\chi(at)\chi(at').
\end{equation}
This integral is of the form $A_1^{(1)}$ in Eq.~(\ref{A1}).  
Therefore one can integrate by parts, twice with respect to $t$ and
twice with respect to $t'$, neglecting the cut-off factor $\chi(at)\chi(at')$.  Then, we find
\beq
\Delta E_2 = \frac{1}{3}\Delta E_1,
\eeq
where $\Delta E_1$ is given by Eq.~(\ref{deltaE1}).  By adding $\Delta E_1$ and $\Delta E_2$,
we find the total correction to the Larmor formula at order $e^2\hbar$ to be
\beq
\Delta E = \frac{e^2 \hbar}{6\pi^2mc^5} \int dt \, dt' \left(\frac{d^3\vect{x}}{dt^3} \cdot \frac{d^2\vect{x}'}{dt'^2} - \frac{d^2\vect{x}}{dt^2} \cdot \frac{d^3\vect{x}'}{dt'^3}\right) \frac{1}{t-t'}. \label{total-time}
\eeq

Let us estimate the size of this correction in a simple situation where the acceleration is linear and 
given by $a(t) = a_0 (1-t^2/t_0^2)$ for $|t| \leq t_0$ and $a(t) = 0$ otherwise. We find
\beq
\Delta E = -\frac{4e^2\hbar{a_0}^2}{3\pi^2mc^5}.
\eeq
On the other hand, the energy of radiation emitted according to the Larmor formula can be found from Eq.~(\ref{emitted}) as $E_{\rm em}^{(0)}= 8{a_0}^2t_0/45\pi c^3$. Hence we have
\beq
\frac{|\Delta E|}{E_{\rm em}^{(0)}} = \frac{15\hbar}{2\pi mc^2t_0}.
\eeq
Therefore, the Larmor formula is expected to be a good approximation as long as $t_0 \gg \hbar/mc^2$, 
which is the time for a light ray to traverse a Compton wavelength of the charged scalar particle.  
Since the probability distribution for the frequency of the photon emitted is given by the square of 
the Fourier transform of $a(t)$, the typical energy of the photon emitted will be
of order $\hbar/t_0$ (though the probability of emission can be made small by letting $a_0$ be small).  This energy 
will be comparable to $mc^2$ if $t_0 \sim \hbar/mc^2$.  Then the scattered charged scalar 
particle will be relativistic, and it is not surprising that the non-relativistic approximation will break down.  It is interesting that the classical (non-relativistic) Larmor formula seems to remain a good approximation 
as long as the scattered state remains non-relativistic even if its momentum may be 
much different from that of the initial state, in the case where the particle is accelerated by a time-dependent but space-independent vector potential.

\section{Quantum correction with space-dependent vector potential}\label{sec:Corrections-space}

In this section we treat 
the case in which the potential varies in a space coordinate, taken to be $z$, but is independent of $t$.  
As in the previous section we
assume further that the external vector potential $V_\mu(z)$ is constant except in the interval $[-Z,Z]$, $Z>0$,
with $V_\mu(z)=0$ for $z < -Z$. We do not assume the constant value of $V_\mu(z)$ for $z>Z$ to be $0$.
We further let $V_z(z) = 0$ by a gauge transformation.
  
The mode functions for the scalar particle can be chosen to be proportional to $\exp[(-ip_0t + i\vect{p}_\perp\cdot \vect{x}_\perp)/\hbar]$ with $\vect{x}_\perp=(x,y)$ and $\vect{p}_\perp = (p_x,p_y)$, where $p_x$ and $p_y$ are the
$x$- and $y$-components of the \emph{contraviariant} vector $\vect{p}$. (Below we also
write the $z$-component of a contravariant vector $\vect{b}$ as $b_z$ in general.) We use the WKB approximation for
the ordinary differential equation which determines the $z$-dependence of the mode functions.  The particle 
(as opposed to anti-particle) solution to the field equation thus obtained that is moving in the positive $z$-direction is
\beq
\Phi_{\vect{p}}(t,\vect{x}) = \sqrt{\frac{p}{\kappa_{\vect{p}}(z)}} \exp{\left[\frac{i}{\hbar} \int_0^z \kappa_{\vect{p}}(\zeta)d\zeta\right]} \exp{\left[\frac{i}{\hbar}\left(\vect{p}_{\perp} \cdot \vect{x}_{\perp} - p_0 t\right)\right]}, \label{WKBspace}
\eeq
where the function analogous to the varying energy $\sigma_\vect{p}(t)$ 
in the time-dependent case is now a varying $z$-component of the momentum,
\beq
\kappa_{\vect{p}}(z) = \sqrt{[p_0 - V_0(z)]^2 - |\vect{p}_\perp - \vect{V}_\perp(z)|^2 - m^2}, \label{zmomentum}
\eeq
and where $p = \sqrt{{p_0}^2 - |\vect{p}_\perp|^2 - m^2}$. 
As in the case with a time-dependent vector potential, it can be shown that higher-order corrections to
Eq.~(\ref{WKBspace}) do not contribute to the energy emitted at order $\hbar$.

The Jacobian determinant defined by Eq.~(\ref{Jacobian}) is 
\beq
\frac{\partial\vect{P}}{\partial\vect{p}} = \frac{dP}{dp},
\eeq
where the derivative of $P=\sqrt{{P_0}^2 - |\vect{P}_\perp|^2 - m^2}$, with $P_0 = p_0 - \hbar k$, $\vect{P}_\perp = \vect{p}_\perp - \hbar\vect{k}_\perp$, is taken with $\vect{p}_\perp$ and $\vect{k}$ fixed.
Hence, the energy emitted
is given, in the limit where the momentum distribution is arbitrarily sharply peaked, by
\beq\label{eq:en-z-dep}
E_{\rm em} = 
- \int \frac{d^3 \vect{k}}{16\pi^3} \frac{P_0}{p_0} \frac{dp}{dP} \A_\mu^*(\vect{p},\vect{k}) \A^\mu(\vect{p},\vect{k}),
\eeq
where $\vect{p}$ is the peak value of the momentum distribution.
Many of the details of the calculation which follows find, as one might expect, direct analogues in the time-dependent case. Although occasional mention will be made of these details, many will be left unremarked.

The formula for the emission amplitude, Eq.~(\ref{eq:em-amp}), remains the same.  After integrating over $t$, $\vect{x}_\perp$ and $\vect{p}'$, we find
\begin{align}
\A_0(\vect{p},\vect{k}) =& \frac{e}{2}\sqrt{\frac{p}{P}} \int dz \, e^{-i k_z z} \frac{1}{\sqrt{\kappa_{\vect{p}}(z)\kappa_{\vect{P}}(z)}}\nonumber\\
	&\times \left[2V_0(z) - (p_0 + P_0)\right] \exp\left\{\frac{i}{\hbar}\int_0^z \left[\kappa_{\vect{p}}(\zeta) - \kappa_{\vect{P}}(\zeta)\right]d\zeta\right\},\\
\A_\perp(\vect{p},\vect{k}) = &\frac{e}{2}\sqrt{\frac{p}{P}} \int dz \, e^{-i k_z z} \frac{1}{\sqrt{\kappa_{\vect{p}'}(z)\kappa_{\vect{p}'}(z)}}\nonumber\\
	&\times \left[2\vect{V}_\perp(z) -(\vect{p}_\perp + \vect{P}_\perp)\right] \exp{\left\{\frac{i}{\hbar} \int_0^z \left[\kappa_{\vect{p}}(\zeta) - \kappa_{\vect{P}}(\zeta)\right]d\zeta\right\}},
\end{align}
and
\begin{align}
\A_z(\vect{p},\vect{k}) =& \frac{ie\hbar}{2}\sqrt{\frac{p}{P}} \int dz \, e^{-i k_z z} \frac{1}{\sqrt{\kappa_{\vect{p}}(z)\kappa_{\vect{P}}(z)}}\nonumber\\
	&\times \left\{-\frac{1}{2}\left[\frac{\kappa_{\vect{p}}'(z)}{\kappa_{\vect{p}}(z)} - \frac{\kappa_{\vect{P}}'(z)}{\kappa_{\vect{P}}(z)}\right] + \frac{i}{\hbar}\left[\kappa_{\vect{p}}(z) + \kappa_{\vect{P}}(z)\right]\right\}\nonumber\\
	& \times \exp{\left\{\frac{i}{\hbar}\int_0^z \left[\kappa_{\vect{p}}(\zeta) - \kappa_{\vect{P}}(\zeta)\right]d\zeta\right\}}.
	\label{lasteq}
\end{align}
Thus, to order $\hbar$, i.e.\ letting $\kappa_\vect{p}'(z)/\kappa_\vect{p}(z) \approx \kappa_\vect{P}'(z)/\kappa_\vect{P}(z)$
in Eq.~(\ref{lasteq}), we have
\begin{align}
\left|\A(\vect{p},\vect{k})\right|^2 = & \frac{e^2}{4}\frac{p}{P} \int dz \, dz' \,  \frac{e^{ik_z(z'-z)}}{\sqrt{\kappa_{\vect{p}}(z)\kappa_{\vect{P}}(z)\kappa_{\vect{p}}(z')\kappa_{\vect{P}}(z')}}\exp{\left\{\frac{i}{\hbar}\int_{z'}^{z}\left[\kappa_{\vect{p}}(\zeta) - \kappa_{\vect{P}}(\zeta)\right]d\zeta\right\}}\nonumber \\
& \times \left\{-\left[2V_0(z) - (p_0 + P_0)\right]\left[2V_0(z') - (p_0 + P_0)\right]\right.\nonumber \\
& \left. 
\,\,\,\,\,\,\,\,+ \left[2\vect{V}_\perp(z) - (\vect{p}_\perp + \vect{P}_\perp)\right]\cdot\left[ 2\vect{V}_\perp(z') - (\vect{p}_\perp + \vect{P}_\perp)\right]\right.\nonumber \\
	& \left.\,\,\,\,\,\,\, 
	+\left[\kappa_{\vect{p}}(z) + \kappa_{\vect{P}}(z)\right]\left[\kappa_{\vect{p}}(z') + \kappa_{\vect{P}}(z')\right]
	\right\}.
	\label{eq:z-dep-sq-em-amp}
\end{align}
We find the energy emitted, $E_{\rm em}$, 
by substituting this formula into Eq.~(\ref{eq:en-z-dep}).  We can simplify $E_{\rm em}$ by
noting that
\beq
\frac{P_0}{p_0}\frac{dp}{dP}\frac{p}{P} = 1,
\eeq
which can readily be proved by using $dP/dP_0 = P_0/P$, $dp/dp_0 = p_0/p$ and $dP_0/dp_0 = 1$.
The following formulas are crucial in expressing the energy emitted in terms of the motion of the corresponding
classical particle:
\beqa
\frac{\vect{p}_\perp - \vect{V}_\perp}{\kappa_{\vect{p}}(z)} & = & \left.\frac{d\vect{x}_\perp}{dz}\right|_\vect{p},\\
\frac{p_0 - V_0}{\kappa_{p}(z)} & = & \left.\frac{dx^0}{dz}\right|_{\vect{p}},
\eeqa
where $x^\mu(z)$ is the world line of the classical particle under the potential $V^\mu(z)$, 
and where `${}|_{\vect{p}}$' indicates that the quantity is evaluated with the initial momentum $\vect{p}$.
We obtain
\begin{multline}
E_{\rm em} = - \frac{e^2}{8} \int \frac{d^3 \vect{k}}{(2\pi)^3} \int dz \, dz' e^{ik_z(z'-z)} \exp{\left\{\frac{i}{\hbar} \int_{z'}^{z} \left[\kappa_{\vect{p}}(\zeta) - \kappa_{\vect{P}}(\zeta)\right]d\zeta\right\}}\\
	\times \left[ \left. \sqrt{\frac{\kappa_{\vect{p}}(z)}{\kappa_{\vect{P}}(z)}}\frac{dx^\mu}{dz} \right|_{\vect{p}} + \left. \sqrt{\frac{\kappa_{\vect{P}}(z)}{\kappa_{\vect{p}(z)}}}\frac{dx^\mu}{dz} \right|_{\vect{P}} \right] \left[ \left. \sqrt{\frac{\kappa_{\vect{p}}(z')}{\kappa_{\vect{P}}(z')}}\frac{dx_\mu'}{dz'} \right|_{\vect{p}} + \left. \sqrt{\frac{\kappa_{\vect{P}}(z')}{\kappa_{\vect{p}(z')}}}\frac{dx_\mu'}{dz'} \right|_{\vect{P}} \right]. \label{EEM}
\end{multline}
The correction to the energy emitted to first order in $\hbar$ again can be attributed to two sources: the
exponential factor and the other multiplicative factor.  We shall examine these separately but combine the intermediate
results to simplify the calculation.

To consider the contribution from the exponential factor, we need to find the expansion of 
$\kappa_{\vect{P}}(z)$ to second order in $\hbar$, which is analogous to that of $\sigma_\vect{P}(t)$ in the 
time-dependent case.  We have
\begin{equation}
\kappa_{\vect{P}}(z) = \kappa_{\vect{p}}(z)\left\{ 1 - \frac{\hbar\omega}{\kappa_\vect{p}}\left(\frac{dt}{dz} - 
\frac{\vect{n}_\perp}{c}\cdot\frac{d\vect{x}_\perp}{dz}\right)
+ \frac{\hbar^2\omega^2}{2\kappa_\vect{p}^2}\left[ - \left( \frac{dt}{dz} - \frac{\vect{n}_\perp}{c}
\cdot	\frac{d\vect{x}_\perp}{dz}\right)^2 + \frac{n_z^2}{c^2}\right]\right\}. \label{kappaP}
\end{equation}
Therefore, the correction to the energy emitted coming from the exponential factor is
\begin{multline}
\Delta E_1 = \frac{ie^2\hbar}{32\pi^3m} \int d\Omega \int_0^\infty 
d\omega\omega^4 \int dt \, dt' e^{i\omega(t-t')-i\omega\vect{n}\cdot(\vect{x}-\vect{x}')/c} 
(\dot{\vect{x}}\cdot\dot{\vect{x}'} - c^2)\\
	\times\int_{t'}^{t} \left(1 - 2 \frac{\vect{n}_{\perp}}{c} \cdot \frac{d\vect{x}_{\perp}}{dT}
	\right) \left(\frac{dz}{dT}\right)^{-2} dT, \label{ill2}
\end{multline}
where $(\vect{x}_\perp(T),z(T))$ is the position of the corresponding classical particle at time $T$ with
$(\vect{x}_\perp(0),z(0)) = (\vect{0},0)$. After inserting the cut-off factor $\chi(at)\chi(at')$, we again find that
the integral is of the form $A_1^{(n)}$, $n=1,2,3$, in Eq.~(\ref{A1}).  Hence, one may integrate the ill-defined integral
(\ref{ill2}) formally by parts.  This means that
\beq
\int_0^\infty 
d\omega\omega^4 \int dt \, dt' e^{i\omega(t-t')} (- c^2)
\int_{t'}^{t} \left(1 - 2 \frac{\vect{n}_{\perp}}{c} \cdot \frac{d\vect{x}_{\perp}}{dT}
	\right) \left(\frac{dz}{dT}\right)^{-2} dT = 0. \label{zerozero}
\eeq
By expanding the factor $e^{-i\omega\vect{n}\cdot(\vect{x}-\vect{x}')/c}$ to second order in $\omega/c$ and integrating over
$\vect{n}$, we obtain to lowest order in $c^{-1}$
\begin{multline}
\Delta E_1 = \frac{ie^2\hbar}{8\pi^2 m} \int_0^\infty d\omega \int dt \, dt' \omega^4
e^{i\omega(t-t'+i\eps)}\chi(at)\chi(at')\\
	\times \left[\left(\dot{\vect{x}}\cdot\dot{\vect{x}'}+\frac{\omega^2}{6}|\vect{x}-\vect{x}'|^2\right)\int_{t'}^{t} \dot{z}^{-2} dT -\frac{2i\omega}{3} (\vect{x}_\perp - \vect{x}_\perp')\cdot \int_{t'}^{t} \dot{\vect{x}}_\perp \dot{z}^{-2} dT\right].
\end{multline}
Again we have replaced $e^{i\omega(t-t')}$ by $e^{i\omega(t-t'+i\eps)}$ to regularize the integral over $\omega$.

We now turn to the contribution from the non-exponential factor in Eq.~(\ref{EEM}).  For $\mu = m \neq 3$, we have
\beq
\left.\frac{dx^m}{dz}\right|_{\vect{P}} = \left.\frac{\kappa_{\vect{p}}(z)}{\kappa_{\vect{P}}(z)} \frac{dx^m}{dz}\right|_{\vect{p}} - \frac{\hbar k^m}{\kappa_{\vect{P}}(z)},
\eeq
where $x^0 \equiv ct$ and $k^0\equiv \omega/c$.
Since we are only looking for corrections at first order in $\hbar$, we find from Eq.~(\ref{kappaP})
\beq
\left.\frac{dx^m}{dz}\right|_{\vect{P}} = \left[1 + \frac{\hbar k_n}{\kappa_{\vect{p}}(z)}\frac{dx^n}{dz}\right] \left.\frac{dx^m}{dz}\right|_{\vect{p}} - \frac{\hbar k^m}{\kappa_{\vect{p}}(z)},
\eeq
where we have used
\beq
k_n \frac{dx^n}{dz} =\omega\left( \frac{dt}{dz} - \frac{\vect{n}_\perp}{c}\cdot \frac{d\vect{x}_\perp}{dz}\right). \label{note}
\eeq
Thus, denoting the contribution from the non-exponential factor by $\Delta E_2$, 
we have the following result, where the summations over Roman indices exclude the 
$z$-component:
\begin{multline}
E^{(0)}_{\rm em} + \Delta E_2 
= -\frac{e^2}{16\pi^3c^3}\int d\Omega \int_0^\infty
 d\omega \omega^2 \int dz \, dz' e^{i\omega(t-t')-i\omega \vect{n}\cdot(\vect{x}-
\vect{x}')/c}\\
	\times \left\{\sqrt{\frac{\kappa_{\vect{p}}(z)}{\kappa_{\vect{P}}(z)}} \sqrt{\frac{\kappa_{\vect{p}}(z')}{\kappa_{\vect{P}}(z')}} \left.\frac{dx^m}{dz}\right|_{\vect{p}} \left.\frac{dx_m'}{dz'}\right|_{\vect{p}} - \frac{\hbar k^m}{2} \left[\frac{1}{\kappa_{\vect{p}}(z')} \left.\frac{dx_m}{dz}\right|_{\vect{p}} + \frac{1}{\kappa_{\vect{p}}(z)} \left.\frac{dx_m'}{dz'}\right|_{\vect{p}}\right] - 1\right\}.
\end{multline}
Therefore, using Eqs.~(\ref{kappaP}) and (\ref{note}), 
we can write
\begin{eqnarray}
\Delta E_2 & = & -\frac{e^2\hbar}{32\pi^3c^3} \int d\Omega \int_0^\infty
 d\omega \omega^2 \int dz \, dz' e^{i\omega(t-t')-i\omega \vect{n}\cdot(\vect{x}-
\vect{x}')/c}\nonumber \\
&& \times \left\{\left[\frac{k_n}{\kappa_{\vect{p}}(z)}\frac{dx^n}{dz} + \frac{k_n}{\kappa_\vect{p}(z')}
\frac{dx^{\prime n}}{dz'}\right]\frac{dx^m}{dz}\frac{dx'_m}{dz'}- \left[\frac{k_n}{\kappa_{\vect{p}}(z')}\frac{dx^n}{dz} + \frac{k_n}{\kappa_{\vect{p}}(z)} \frac{dx^{\prime n}}{dz'}\right]\right\}.\nonumber \\
\end{eqnarray}
Collecting only the terms up to order $c^0$ in the integrand, we have
\begin{eqnarray}
\Delta E_2 & = & -\frac{e^2\hbar}{32\pi^3c^3} \int d\Omega \int_0^\infty d\omega \omega^3 \int dt \, dt' e^{i\omega(t-t')-i\omega \vect{n}\cdot(\vect{x}-
\vect{x}')/c}\nonumber \\
&& \times \left\{
\left[\frac{1}{\dot{z}\kappa_{\vect{p}}(z)}\left( 1 - \frac{\vect{n}_\perp}{c}\cdot \dot{\vect{x}}_\perp\right) +
\frac{1}{\dot{z}'\kappa_{\vect{p}}(z')}\left( 1 - \frac{\vect{n}_\perp}{c}\cdot \dot{\vect{x}'}_\perp\right)\right]
(c^2 - \dot{\vect{x}}_\perp\cdot \dot{\vect{x}}'_\perp) - \frac{2}{m} \right\}, \nonumber\\
\label{use}
\end{eqnarray}
where we have used $\kappa_\vect{p}(z) \approx mdz/dt$ to lowest order in $c^{-1}$.  
The argument that led to Eq.~(\ref{zerozero}) can be used to conclude that
\beq
c^2\int_0^\infty d\omega \omega^3 \int dt \, dt' e^{i\omega(t-t')}
\left[\frac{1}{\dot{z}\kappa_{\vect{p}}(z)} +
\frac{1}{\dot{z}'\kappa_{\vect{p}}(z')}\right]=0.
\eeq
Expanding the factor
$e^{-i\omega\vect{n}\cdot(\vect{x}-\vect{x}')/c}$ to order $\omega^2/c^2$ and carrying out the $\vect{n}$-integration, we find
\begin{multline}
\Delta E_2 = \frac{e^2\hbar}{8\pi^2mc^3} \int_0^\infty d\omega \omega^3 \int dt \, dt' e^{i\omega(t-t'+i\eps)}\chi(at)\chi(at') \\
	\times \left[\left(\dot{\vect{x}}_\perp\cdot\dot{\vect{x}}'_\perp + \frac{\omega^2}{6}|\vect{x}-\vect{x}'|^2\right)
	(\dot{z}^{-2} + \dot{z}'^{-2}) - \frac{i\omega}{3} (\vect{x}_\perp - \vect{x}_\perp')\cdot(\dot{\vect{x}}_\perp \dot{z}^{-2} + \dot{\vect{x}}_\perp' \dot{z}'^{-2})\right].
\end{multline}

It is convenient to combine the two integrals $\Delta E_1$ and $\Delta E_2$ at this point.  After integrating over $\omega$,
$\Delta E \equiv \Delta E_1 + \Delta E_2$ to order $c^{-3}$ becomes
\begin{equation}
\Delta E = \frac{e^2 \hbar}{8 \pi^2 mc^3} \int dt \, dt' \chi(at) \chi(at')
\left[ F_1(t,t') + F_2(t,t') + F_3(t,t')\right], \label{DeltaE}
\end{equation}
where
\begin{eqnarray}
F_1(t,t') & \equiv & 
-\frac{4!\dot{\vect{x}}\cdot\dot{\vect{x}}'}{(t-t'+i\eps)^5} \int_{t'}^{t} \dot{z}^{-2} dT  
+ \frac{3!\dot{\vect{x}}_\perp \cdot \dot{\vect{x}}_\perp' \left(\dot{z}^{-2} + \dot{z}'^{-2}\right)}{(t-t'+i\eps)^4},
\label{F1}\\
F_2(t,t') & \equiv &
	\frac{1}{6}\left[\frac{6!|\vect{x}-\vect{x}'|^2}{(t-t'+i\eps)^7}\int_{t'}^{t} \dot{z}^{-2} dT - 
	\frac{5!|\vect{x}-\vect{x}'|^2(\dot{z}^{-2} + \dot{z}'^{-2})}{(t-t'+i\eps)^6}\right],\label{F2}\\
F_3(t,t') & \equiv &	
\frac{1}{3}\left[\frac{2\cdot 5!(\vect{x}_\perp - \vect{x}_\perp')}{(t-t'+i\eps)^6}\cdot\int_{t'}^{t} \dot{\vect{x}}_\perp \dot{z}^{-2} dT +\frac{4!(\vect{x}_\perp -\vect{x}_\perp')\cdot(\dot{\vect{x}}_\perp \dot{z}^{-2} + \dot{\vect{x}}_\perp' \dot{z}'^{-2})}{(t-t'+i\eps)^5}\right].
\nonumber \\ \label{F3}
\end{eqnarray}
Note that all terms in Eq.~(\ref{DeltaE}) are of the form $A_1^{(n)}$ in Eq.~(\ref{A1}).
For example, the integral in Eq.~(\ref{DeltaE}) involving the first term of $F_2(\omega,t,t')$  can be seen to be of the
form $A_1^{(3)}$ with $f(t,t') = 1$, $g_1(t) = g_2(t)=x_i(t)$ and
$g_3(t) = \int_0^t \dot{z}^{-2}dT$.  Thus, we integrate by parts to reduce the denominator to  $(t-t'+i\eps)^3$
in each term in Eqs.~(\ref{F1})--(\ref{F3}).  We integrate the terms proportional to $\int_{t'}^t \dot{z}^{-2} dT$ so that
the coefficient functions are differentiated with respect to each of $t$ and $t'$ twice, and for the rest
we choose to integrate by parts so that there is no derivative on either $\dot{z}^{-2}$ or $\dot{z}^{\prime -2}$.
Thus, we find
\beqa
F_1(t,t') & \sim & \frac{2}{(t-t'+i\eps)^3}\left(\ddot{\vect{x}}\cdot\ddot{\vect{x}}'\int_{t'}^t \dot{z}^{-2}dT
-  \ddot{z}\dot{z}^{\prime -1}+ \ddot{z}'\dot{z}^{-1}\right),\\
F_2(t,t') & \sim & \frac{2}{3(t-t'+i\eps)^3}\left( - \ddot{\vect{x}}\cdot\ddot{\vect{x}}'\int_{t'}^t \dot{z}^{-2}dT
+ \ddot{\vect{x}}\cdot\dot{\vect{x}}'\dot{z}^{\prime -2} - \ddot{\vect{x}}'\cdot\dot{\vect{x}}\dot{z}^{-2}\right),\\
F_3(t,t') & \sim & -\frac{2}{3(t-t'+i\eps)^3}\left( \ddot{\vect{x}}_\perp\cdot\dot{\vect{x}}_\perp'\dot{z}^{\prime -2}
-\ddot{\vect{x}}_\perp'\cdot \dot{\vect{x}}_\perp\dot{z}^{-2}\right),
\eeqa
where $\sim$ indicates equivalence under integration over $t$ and $t'$.  Adding these three terms and integrating some terms
further by parts, we find 
\beq
\Delta E = \frac{e^2 \hbar}{12 \pi^2 m c^3} \int dt \, dt' \left[
\frac{2 \ddot{\vect{x}}_\perp \cdot \ddot{\vect{x}}_\perp'}{(t-t'+i\eps)^3}- \frac{\dddot{z} \, \dddot{z}'}{t-t'+i\eps} \right] \int_{t'}^{t} \dot{z}^{-2} dT.
\eeq
A form more convenient for concrete calculations can be found by integrating the first term by parts further as
\begin{equation}
\Delta E = \frac{e^2 \hbar}{12 \pi^2 m c^3} \int \frac{dt \, dt'}{t-t'} \left(-\dddot{\vect{x}} \cdot \dddot{\vect{x}}' \int_{t'}^{t} \dot{z}^{-2} dT +\dddot{\vect{x}}_\perp\cdot \ddot{\vect{x}}'_\perp \dot{z}^{\prime -2}
- \dddot{\vect{x}}'_\perp \cdot \ddot{\vect{x}}_\perp\dot{z}^{-2}\right). \label{total-space}
\end{equation}
This correction is of the same order in $c^{-1}$ as the Larmor formula, though it is of course of higher order in $\hbar$, 
in contrast to the correction (\ref{total-time}) for a time-dependent vector potential, which is of higher order in $c^{-1}$.

To estimate the size of this correction, we consider a charged particle moving at a constant speed $v_z$ in the $z$-direction and accelerated in the $x$-direction with acceleration given by $a(t)=a_0(1-t^2/t_0^2)$ 
for $|t| \leq t_0$.  It is possible to arrange
the vector potential to realize this motion as shown in Appendix~\ref{app:motion}.  The first term in brackets in 
Eq.~(\ref{total-space}) gives a vanishing contribution.  {}From the remaining terms we find
\beq
\Delta E = -\frac{2e^2\hbar{a_0}^2}{3\pi^2mv_z^2 c^3},
\eeq
and
\beq
\frac{|\Delta E|}{E_{\rm em}^{(0)}} = \frac{15\hbar}{4\pi mv_z^2t_0},
\eeq
where $E_{\rm em}^{(0)} = 8{a_0}^2t_0/45\pi c^3$ is the energy emitted according to the Larmor formula as before.
Thus, the correction is small and expected to be reliable 
as long as the kinetic energy associated with the motion in the $z$-direction is much larger
than the typical energy of the photon emitted, $\hbar/t_0$.

\section{Summary and outlook}\label{sec:Conclusions}

In this paper we showed that the energy and momentum of radiation emitted by a charged scalar particle in QED agree with
the classical result (\ref{Larmor4}) at order $e^2$ in the limit $\hbar\to 0$ and 
then went on to study the correction (at first order in $\hbar$) to the energy emitted in the non-relativistic limit in two cases: one with a time-dependent but space-independent vector potential and the other with a time-independent 
vector potential which depends on one space coordinate, $z$. Both corrections were found to arise entirely due to
the fact that the momenta of the initial and final scalar wave functions are different
in the emission amplitude.   The results are given by
Eqs.~(\ref{total-time}) and (\ref{total-space}).  They are expressed in terms of the classical trajectory and 
are different from each other.  Thus, the quantum correction is sensitive to how the particle is accelerated
as well as to the motion of the corresponding classical particle.
Another notable feature of these corrections 
is that they are non-local in time unlike the classical approximation.

We estimated the size of the correction in each case for a given acceleration of simple form. 
For the time-dependent potential the correction is much smaller than the classical result unless the typical energy of the
photon emitted is comparable to the rest mass energy of the particle, 
with the non-relativistic approximation itself breaking down.
On the other hand, for the $z$-dependent potential,
the correction is small compared to the classical result if the typical energy of the photon emitted is much smaller than the kinetic energy of the particle in the $z$-direction. 

It would be interesting to test the quantum corrections to the Larmor formula obtained in this paper though it would be
difficult to realize the conditions in which our results can directly be compared with experimental results.
One quantum system to which the Larmor formula and other classical results are applicable is a Rydberg atom, i.e.\ an atom
with an electron with a very high principal quantum number up to a few hundred.   Indeed the Larmor formula is known to give
a very good approximation to the life time of states with high principal and angular-momentum quantum 
numbers~\cite{PhysRevA1,PhysRevA2}.  It would be
interesting to calculate the quantum correction to this approximation by extending our calculations to cases with a charged particle in a radially varying potential and possibly to extend our results to the case with a vector potential varying in a more general way.

\acknowledgments

We thank Kazuhiro Yamamoto for useful correspondence which motivated this work.

\appendix

\section{Cut-off independence of integrals in Secs.~\ref{sec:Corrections-time} and \ref{sec:Corrections-space}}\label{app:cut-off}

In this Appendix we show that formal integration by parts used in this paper to find the quantum corrections is justified.
Let $I \equiv [-T,T]$, $T > 0$.  We choose this interval such that the acceleration of the particle is nonzero only
if $t\in I$. Let $f(t,t')$ be a smooth function such that 
the support of $\partial_t f(t,t')$ (resp.~$\partial_{t'}f(t,t')$) is a subset of $I \times \mathbb{R}$ 
(resp.~$\mathbb{R}\times I$).  Then, one can show that $f$, $\partial_t f$ and $\partial_{t'}f$ are all bounded.
Let $g_i(t)$, $i=1,2,\ldots, n$, be smooth functions such that the support of $g_i''(t)$ is a subset of $I$.
Then, it can readily be seen that $g_i'$ are bounded.
We let $\chi(t)$ be a smooth function such that it is compactly supported with $\chi(t)=1$ for $t\in I$.  We use $\chi(at)$, $0<a\leq 1$, as our cut-off factor, with the limit $a\to 0$ taken at the end.  Note that
$\lim_{a\to 0}a^2\int_{-\infty}^\infty \left[\chi'(at)\right]^2 dt = 0$.
This property was necessary for the cut-off factor for deriving the Larmor formula in Sec.~\ref{sec:Larmor}.
All integrals in Secs.~\ref{sec:Corrections-time} and \ref{sec:Corrections-space} 
that are ill-defined without the cut-off factor take the form
\begin{equation}
A_1^{(n)}  =  \int dtdt' \frac{f(t,t')}{(t-t'+i\eps)^{n+4}}
\left\{\prod_{i=1}^n\left[g_i(t)-g_i(t')\right]\right\}\chi(at)\chi(at'). \label{A1}
\end{equation}
What we show in this Appendix is that this integral can be reduced to the sum of integrals with no derivatives on the cut-off
factor and convergent without them and those which tend to zero as $a\to 0$.  This implies that one can use formal integration
by parts for this integral until it is convergent, as we did in Secs.~\ref{sec:Corrections-time} and \ref{sec:Corrections-space}.

We first prove that the integral of the following form is convergent without the cut-off factor:
\begin{equation}
A_2^{(n)}  =  \int dtdt' \frac{\partial_t f(t,t')}{(t-t'+i\eps)^{n+3}}
\left\{\prod_{i=1}^n\left[g_i(t)-g_i(t')\right]\right\}\chi(at)\chi(at').
\end{equation}
Since $\partial_t f(t,t')=0$ for $t \notin I$, the $t$-integral is effectively over
interval $I$.  We have noted that $\partial_tf(t,t')$ is bounded.  Since $g_i''(t)$ are nonzero only for $t\in I$, we have
$g_i(t')=\alpha_i^-t' + \beta_i^-$ for $t' < -T$ and $g_i(t')=\alpha_i^+t' + \beta_i^+$ for $t' > T$ for some constants
$\alpha_i^\pm$ and $\beta_i^\pm$.  Then it is clear that the $t'$-integral is convergent without the cut-off factor.

Next we prove that the following integral tends to zero as $a\to 0$:
\begin{equation}
A_3^{(n)}  =  a\int dtdt' \frac{f(t,t')}{(t-t'+i\eps)^{n+3}}\left\{\prod_{i=1}^n\left[g_i(t)-g_i(t')\right]\right\}
\chi'(at)\chi(at').
\end{equation}
To this end it is useful to prove that the following integrals tend to zero as $a\to 0$:
\begin{eqnarray}
A_4^{(n)} & = & a\int dtdt' \frac{\partial_{t'}f(t,t')}{(t-t'+i\eps)^{n+2}}\left\{\prod_{i=1}^n\left[g_i(t)-g_i(t')\right]\right\}
\chi'(at)\chi(at'), \label{A4}\\
A_5^{(n)} & = &  a^2\int dtdt' \frac{f(t,t')}{(t-t'+i\eps)^{n+2}}\left\{\prod_{i=1}^n\left[g_i(t)-g_i(t')\right]\right\}
\chi'(at)\chi'(at').
\end{eqnarray}
The $t'$-integral in Eq.~(\ref{A4}) 
is over the interval $I$, and hence we can drop the cut-off factor $\chi(at')$.  
Furthermore, $\partial_{t'}f(t,t') \equiv F(t')$ 
is $t$-independent where $\chi'(at)\neq 0$.  Then, by letting $t=\eta/a$, we have
\beqa
A_4^{(n)} & = & a^2 \int_{-\infty}^{-T} d\eta \int_{-T}^{T}dt'
\frac{F(t')}{(\eta-at')^{n+2}}\left\{\prod_{i=1}^n \left[ \alpha_i^-\eta + a(\beta_i^- - g_i(t'))\right]\right\}
\chi'(\eta) \nonumber \\
&& + a^2 \int^{\infty}_{T} d\eta \int_{-T}^{T}dt'
\frac{F(t')}{(\eta-at')^{n+2}}\left\{\prod_{i=1}^n \left[ \alpha_i^+\eta + a(\beta_i^+ - g_i(t'))\right]\right\}
\chi'(\eta).
\eeqa
These integrals have finite limits as $a\to 0$.  Hence, $A_4^{(n)}\to 0$ as $a\to 0$.  To show that $A_5^{(n)}\to 0$ as
$a\to 0$, we note that $f(t,t')$ is constant if $|t|, |t'|>T$.  The integral $A_5^{(n)}$ has nonzero contributions only from the four disjoint regions $(T,\infty)\times (T,\infty)$, $(-\infty,-T)\times (T,\infty)$,
$(-\infty,-T)\times (-\infty,-T)$ and $(T,\infty)\times (-\infty,-T)$ on the $tt'$-plane because of the factor
$\chi'(at)\chi'(at')$. 
Let $f(t,t')=f_{++}$ in the first region.  
Then the contribution from the first region to $A_5^{(n)}$ can be written, after the
change of variables $t=\eta/a$, $t'=\eta'/a$,
\beq
A_5^{(n)}|_{++} = a^2f_{++}\prod_{i=1}^n \alpha_i^{+}\int_{T}^\infty d\eta \int_{T}^\infty d\eta' \frac{\chi'(\eta)\chi'(\eta')}{(\eta-\eta'+i\eps)^{2}}.
\eeq
The contribution from $(-\infty,-T)\times (-\infty,-T)$ has a similar expression.
The contribution from $(T,\infty)\times (-\infty,-T)$ with $f(t,t')=f_{+-}$ is
\beq
A_5^{(n)}|_{\pm\mp} = a^2f_{+-}\int_{T}^\infty d\eta \int_{-\infty}^{-T} d\eta'
\frac{\chi'(\eta)\chi'(\eta')}{(\eta-\eta')^{n+2}}\prod_{i=1}^n
\left(\alpha_i^+\eta + a\beta_i^+ - \alpha_i^-\eta' - a\beta_i^-\right),
\eeq
and that from $(-\infty,-T)\times (T,\infty)$ is similar.  Hence we find that $A_5^{(n)}\to 0$ as $a\to 0$.

To show that $A_3^{(n)}\to 0$ as $a\to 0$, we first integrate by parts and find
\begin{align}
A_3^{(n)}
	&= -\frac{a}{n+2} \int dt \, dt' \, \frac{1}{(t-t'+i\eps)^{n+2}} \frac{\partial}{\partial t'}\left\{f(t,t') \prod_{i=1}^n \left[g_i(t)-g_i(t')\right] \chi(at')\right\}\chi'(at) \nonumber \\
	&= -\frac{A_4^{(n)}}{n+2} - \frac{A_5^{(n)}}{n+2}\nonumber \\
	&\qquad + \frac{a}{n+2} \sum_{k=1}^n
	\int dt \, dt' \frac{f(t,t')g_k'(t')}{(t-t'+i\eps)^{n+2}}\left\{\prod_{i \neq k} \left[g_i(t)-g_i(t')\right]
	\right\} \chi'(at) \chi(at').
\end{align}
Now, the first and second terms tend to zero as $a\to 0$.  Each of the remaining terms is of the form $A_3^{(n-1)}$ because
the partial derivatives of $f(t,t')g_k'(t')$ with respect to $t$ and $t'$ have support in $I\times \mathbb{R}$ and
$\mathbb{R}\times I$, respectively.  
Therefore, if $A_3^{(n-1)}$ tends to zero as $a\to 0$, so does $A^{(n)}_3$.  This means that 
all we need to show is $A_3^{(0)}\to 0$ as $a\to 0$, which is true because
\begin{equation}
A_3^{(0)} = -\frac{1}{2} A_4^{(0)} - \frac{1}{2} A_5^{(0)}.
\end{equation}

Now, we are ready to turn to the integrals $A_1^{(n)}$, which are the ones we encounter in our calculations.  By
integrating by parts with respect to $t$ we find
\begin{eqnarray}
A_1^{(n)}
	& = & \frac{1}{n+3} \int dt \, dt' \frac{1}{(t-t'+i\eps)^{n+3}} \frac{\partial}{\partial t} \left\{f(t,t') \prod_{i=1}^n \left[g_i(t)-g_i(t')\right] \chi(at)\right\} \chi(at') \nonumber \\
 &	= & \frac{1}{n+3} \left[A_2^{(n)} + A_3^{(n)}\right. \nonumber \\
&& \left. + \sum_{k=1}^n\int dt \, dt'
\frac{f(t,t')g_k'(t) }{(t-t'+i\eps)^{n+3}}\left\{\prod_{i \neq k}\left[g_i(t)-g_i(t')\right]\right\} \chi(at) \chi(at')\right].
\end{eqnarray}
As we have seen, the term $A_2^{(n)}$ is cut-off independent and $A_3^{(n)}\to 0$ as $a\to 0$.  The remaining terms are
of the form $A_1^{(n-1)}$.  Thus, all we need to show is that $A_1^{(0)}$ can written in a cut-off independent form.  Indeed
we have
\begin{equation}
A_1^{(0)} =  \frac{1}{3} A_2^{(0)} + \frac{1}{3} A_3^{(0)} \to \frac{1}{3}A_2^{(0)}\,\,{\rm as}\,\, a\to 0.
\end{equation}

Thus, we have shown that we may use integration by parts with respect to $t$ for integrals of the form $A_1^{(n)}$ 
disregarding the cut-off factor until it is convergent without them.  
It is clear that this statement holds for integration by parts with respect to $t'$ as well.

\section{The vector potential for the motion used in Sec.~\ref{sec:Corrections-space}}\label{app:motion}

We recall that the local momentum of the particle is given by
\beqa
m\frac{d\vect{x}_\perp}{d\tau} & = & \vect{p}_\perp - \vect{V}_\perp(z),\\
m\frac{dz}{d\tau} & = & \sqrt{\left[p_0-V_0(z)/c\right]^2 - |\vect{p}_\perp - \vect{V}_\perp(z)|^2 - m^2c^2},
\eeqa
where we have restored the factors of $c$ in  Eq.~(\ref{zmomentum}) letting $V_0$ and $\vect{V}_\perp$ have
the dimensions of energy and momentum, respectively.  Here $\vect{p}_\perp$ and $p_0$ are constants.  Thus,
it is clear that any motion in the perpendicular direction can be realized by adjusting $\vect{V}_\perp(z)$ appropriately 
while maintaining the condition $dz/dt \approx dz/d\tau = v_z$ by adjusting $V_0(z)$.

\end{document}